\documentclass[fleqn,twoside]{article}
\usepackage{espcrc2}

\usepackage{graphicx}
\usepackage[figuresright]{rotating}

\newcommand{\AmS}{{\protect\the\textfont2
  A\kern-.1667em\lower.5ex\hbox{M}\kern-.125emS}}
\title{Dark energy and the hierarchy problem}

\author{Pisin Chen\address[MCSD]{Kavli Institute for Particle Astrophysics and
Cosmology, \\
Stanford Linear Accelerator Center, Stanford University, Stanford CA 94309, U.S.A.}%
        \thanks{This work is supported by the US Department of Energy under
        Contract No.\ DE-AC03-76SF00515.}}

\begin{document}

\begin{abstract}
The well-known hierarchy between the Planck scale ($\sim
10^{19}$GeV) and the TeV scale, namely a ratio of $\sim 10^{16}$
between the two, is coincidentally repeated in a inverted order
between the TeV scale and the dark energy scale at $\sim
10^{-3}$eV implied by the observations. We argue that this is not
a numerical coincidence. The same brane-world setups to address
the first hierarchy problem may also in principle address this
second hierarchy issue. Specifically, we consider supersymmetry in
the bulk and its breaking on the brane and resort to the Casimir
energy induced by the bulk graviton-gravitino mass-shift on the
brane as the dark energy. For the ADD model we found that our
notion is sensible only if the number of extra dimension $n=2$. We
extend our study to the Randall-Sundrum model. Invoking the
chirality-flip on the boundaries for SUSY-breaking, the zero-mode
gravitino contribution to the Casimir energy does give rise to the
double hierarchy. Unfortunately since the higher Kaluza-Klein
modes acquire relative mass-shifts at the TeV level, the zero-mode
contribution to Casimir energy is overshadowed.

\vspace{1pc}
\end{abstract}

\maketitle

\section{INTRODUCTION}

The combination of recent data from WMAP (WMAP3)\cite{WMAP} and
Sloan Digital Sky Survey (SDSS)\cite{Tegmark:2006} determines
$w=-0.94\pm 0.09$ for the dark energy equation of state,
$p=w\rho$, while the Supernova Legacy Survey (SNLS)\cite{SNLS}
gives $w=-1.023\pm 0.090({\rm stat})\pm 0.054({\rm sys})$. These
results do not seem to leave much freedom for ideas other than a
bone fide cosmological constant ($w=-1$). If dark energy is indeed
a cosmological constant which never changes in space and time,
then it must be a fundamental property of the spacetime.
Observations imply that such a cosmological constant would
correspond to a vacuum energy density $\rho_{\rm DE}\sim(10^{-3}
{\rm eV})^4$. One cannot but note that the energy scale of this
fundamental property of the vacuum is so much smaller than that of
the standard model of particle physics, which is $\sim$TeV, by a
factor $\sim 10^{-15}$. Why is this energy gap so huge?

There has been another long-standing hierarchy problem in physics,
i.e., the existence of a huge gap between the Planck scale of
quantum gravity at $10^{19}$GeV and that of the standard model
gauge interactions at TeV, by a factor $\sim 10^{16}$. As is
well-known, there have been two interesting solutions to this
hierarchy problem proposed in recent years: the
Arkani-Hamed-Dimopoulos-Dvali (ADD) model\cite{ADD} and the
Randall-Sundrum (RS) model\cite{RS}. In both models the
brane-world scenario is invoked where the 3-brane is imbedded in
the extra dimensions, and the SM fields are confined to the brane
while gravity fields reside in the bulk. In the case of ADD, the
extra spatial dimensions are flat. The gravity is weak (or the
Planck scale is huge) because it is diluted by the largeness of
the extra dimensions in which it resides. In the alternative
solution proposed by RS, the gravity is weak on the TeV brane
because its strength is exponentially suppressed by the warp
factor descended from the Planck brane.

Are these two hierarchy problems related? To address that, we
notice the remarkable numerical coincidence between the two, i.e.,
\begin{equation}
\frac{\rho_{\rm DE}^{1/4}}{M_{\rm SM}}\simeq \frac{M_{\rm
SM}}{M_{\rm Pl}}.
\end{equation}
That is, the dark energy scale happens to be inverted from the
Planck scale through the intermediary of the Standard Model (SM)
scale, presumably $\sim$ TeV. We suspect that this is not a pure
coincidence. Rather, there may actually be a deeper underlying
connection that results in such a manifestation. This may not be
too much of a stretch. After all, the cosmological constant as a
manifestation of vacuum energy is necessarily connected with the
structure of spacetime and quantum fields. But Eq.(1) actually
implies more. It suggests that their possible connection must be
mediated and inverted by the TeV physics. In this paper we will
rely on Eq.(1) as our guidance to explore the nature of the dark
energy. We should like to caution, however, that the dark energy
scale is actually not on the same footing as the other two energy
scales. Whereas TeV scale represents the interaction strengths of
the Standard Model and possibly its supersymmetric extension, and
the Planck scale that for the gravitational interaction, the dark
energy scale is not associated with any new interaction strength
per se. After all, there are only four fundamental interactions in
this world. It is therefore clear that the dark energy scale must
not be a primary fundamental scale in physics, but rather a
deduced, secondary quantity. In this regard, Eq.(1) serves to
explicate the relationship of the dark energy scale with that of
the four fundamental interactions. That is, we posit that the
underlying mechanism that induces the dark energy must be resulted
from a double suppression by the same hierarchy factor descended
from the Planck scale:
\begin{equation}
\rho_{\rm DE}^{1/4}\simeq \frac{M_{\rm SM}}{M_{\rm pl}}M_{\rm
SM}=\Big(\frac{M_{\rm SM}}{M_{\rm Pl}}\Big)^2 M_{\rm Pl}.
\end{equation}

In our previous attempt, we investigated the Casimir energy on a
supersymmetry-breaking brane as dark energy based on the ADD-like
geometry \cite{Chen:2004fn,Chen:2004zv}. But instead of invoking
Eq.(1) as our guidance, we looked for the general constraint on
the various fundamental energy scales if the Casimir energy so
induced was interpreted as the dark energy. In Sec.2 we will first
correct a mistake in our previous work due to an incorrect
normalization convention. We will show, under the corrected
expression, that the ADD-required extra-dimension size would
provide the right scaling for Casimir energy as dark energy if and
only if the number of extra dimension $n=2$. We will comment on
the interesting implication of this conclusion.

In Section 3 we will apply the same guiding principle to the RS
warped geometry. We invoke the very interesting work by Gherghetta
and Pomarol (PG)\cite{Gherghetta:2001}, where SUSY-breaking in the
RS model is induced not by dynamics, but by a clever choice of
different boundary conditions between the bosonic and the
fermionic bulk fields. There the SUSY-breaking induced zero-mode
gravitino mass is precisely doubly suppressed from the Planck
scale. Unfortunately the higher modes in the graviton-gravitino
Kalaza-Klein tower in this case are at the TeV scale. These higher
modes would bring the resultant Casimir energy back to the TeV
scale, and therefore their contributions would overwhelm that from
the zero-mode. Thus the PG construction is unfortunately not
compatible with our approach to the dark
energy problem. Comments and remarks on possible remedies are given at the end.\\

\section{CASIMIR ENERGY IN ADD GEOMETRY}

Casimir effect has been considered as a possible origin for the
dark energy by many authors
\cite{Milton:2002hx,Gupta:2002,Bauer:2003,Aghababaie:2003wz,Burgess:2004kd,Chen:2004fn,Chen:2004zv}.
It is known that the conventional Casimir energy in the ordinary
3+1 dimensional spacetime cannot provide repulsive gravity
necessary for dark energy. Conversely, Casimir energy on a 3-brane
imbedded in a higher-dimensional world with suitable boundary
conditions can in principle give rise to a positive cosmological
constant. Typically, the resulting Casimir energy density on the
3-brane scales as
\begin{equation}
\rho^{(4)}_{\rm Casimir} \sim a^{-4},
\end{equation}
where $a$ is the extra dimension size. As summarized by
Milton\cite{Milton:2002hx}, the required extra dimension sizes for
it to conform with the supposed dark energy would have to be very
large, roughly consistent with that required for the ADD solution
to the Planck-SM hierarchy.

The scaling of the Casimir energy is modified if the system is
supersymmetric in the bulk but broken on the brane. In addition to
the extra dimension size, the mass-shift of the bulk field under
SUSY-breaking also plays a role. Consider a (3+$n$+1)-dimensional
space-time with $n$ compact extra dimensions of size $a$, in which
the standard model fields and their superpartners are localized on
a 3-brane while the gravity sector resides in the bulk. We assume
that SUSY is preserved in the bulk and only broken on the 3-brane
with a breaking scale $M_{\rm SUSY}$. In this scenario and at tree
level the graviton remains massless everywhere while the gravitino
acquires a mass, $m_{3/2}$, only on the brane. That is, the
gravitino mass is a delta-function along the extra dimension and
localized on the brane.

For the ADD model the compactified $n$ extra dimensions can be
treated as an $n$-torus, $\mathcal{T}^n$. In $\mathcal{M}^4 \times
\mathcal{T}^n$ the renormalized Casimir energy density
$\rho_{\textrm{v}}^{\textrm{(ren)}}(m^2,a)$ is the difference of
the vacuum energy densities at $a$ and infinity:
\begin{equation}
\rho_{\textrm{v}}^{\textrm{(ren)}}(m^2,a) = \rho_{\textrm{v}}
(m^2,a) - \rho_{\textrm{v}} (m^2,a \rightarrow \infty) \, .
\end{equation}
In our SUSY setup and up to $\mathcal{O}\left(m^2\phi^2\right)$,
the 4D Casimir energy density is the sum of that contributed from
graviton and gravitino modes, with the extra dimensions integrated
out\cite{Chen:2004fn}. This gives
\begin{eqnarray}
\rho^{(4)}_{\rm Casimir} & \equiv & \rho_{\rm
gravitino}^{\textrm{(ren)}}(m_{3/2}^2,a) +
\rho_{\rm graviton}^{\textrm{(ren)}} (0,a) \nonumber \\
& \cong & C_n  \cdot a^{-2} \cdot  m_{3/2}^2 \, .
\end{eqnarray}
The coefficient $C_n$ is equal in magnitude but opposite in signs
for graviton and gravitino modes, and is in general insensitive to
the extra dimensionality $n$.

In our previous work the normalization of the $(4+n)$D mass
function was incorrectly treated. As a result, the Casimir energy
on the brane had an extra suppression factor $(\delta/a)^n$, where
$\delta$ is the ``thickness" of the brane where SUSY is broken,
which we assumed to be the string length. As a consequence
$(\delta/a)^n$ provided a huge suppression of the Casimir energy.
We now recognize that this suppression factor was spurious. If the
mass function was properly normalized, this suppression factor
should have disappeared.

As we are motivated to connect the two hierarchies, we invoke the
ADD relation for the large extra dimension size where the
effective gravity strength on the brane is identified with the SM,
or TeV, scale:
\begin{equation}
a\sim M_{\rm Pl}^{2/n}M_{\rm SM}^{-(n+2)/n} \, .
\end{equation}
On the ground of dimensional arguments, the gravitino mass is
related to the SUSY-breaking scale as
\begin{equation}
m_{3/2}\sim \frac{M_{\rm SUSY}^2}{M_{\rm Pl}} \, .
\end{equation}
 Inserting these into Eq.(5), we find
\begin{equation}
\rho^{(4)}_{\rm Casimir} \sim \left( \frac{M_{\rm SUSY}}{M_{\rm
Pl}} \right)^4 \left( \frac{M_{\rm SM}}{M_{\rm Pl}}
\right)^{4/n-2} M_{\rm SM}^4 \, .
\end{equation}
On the other hand, Eq.(1) gives
\begin{equation}
 \rho_{\rm DE}
\sim \left( \frac{M_{\rm SM}}{M_{\rm Pl}} \right)^4 M_{\rm SM}^4\,
.
\end{equation}
Following the spirit of Eq.(1), where there are only three energy
scales in the world, we assume that the SUSY-breaking scale is
also around TeV: $M_{\rm SUSY}\sim M_{\rm SM}\sim{\rm TeV}$. If we
identify the Casimir energy as the dark energy, i.e., that we
equate Eq.(8) and Eq.(9), then our notion would true only if the
number of extra dimension $n=2$. This is remarkable and some
comments are in order.

We note that there is a fundamental difference between such a
Casimir energy and the conventional vacuum energy. The Casimir
energy is nontrivial only around the 3-brane, and it entails the
equations of state: $p_a=-\rho$ and $p_b > 0$, where $p_a$ and
$p_b$ are its pressures along the 3-brane and the extra
dimensions, respectively. In contrast, the brane tension from the
conventional vacuum energy obeys the following equations of state
instead: $p_a=-\rho$ and $p_b = 0$. Study shows that the brane
tension can be perfectly cancelled with the curvature via the
self-tuning mechanism if the number of extra dimensions is
precisely two\cite{Chen:2000at}. On the other hand the Casimir
energy cannot be removed by the same self-tuning procedure and
should survive as the leading contribution to vacuum energy on the
brane. We see that our conclusion of $n=2$ coincides with that
required by the self-tuning mechanism. Therefore $n=2$ is not only
a necessary but also a sufficient condition in order for the
Casimir energy to behave as dark energy, both qualitatively and
quantitatively.

It is interesting to note that a similar concept of SUSY-breaking
in a brane-world, called supersymmetric large extra dimensions
(SLED), has been invoked to address the vacuum energy and the
cosmological constant problem
\cite{Aghababaie:2003wz,Burgess:2004kd}. In that proposal SUSY in
the bulk is broken by the presence of non-supersymmetric 3-branes
imbedded in two extra dimensions. Here we arrive at the same
conclusion on the $n=2$ extra dimensionality, but through a
different route. We note that the same conclusion was reached by
Gupta\cite{Gupta:2002} through the study of the KK modes
contribution to the vacuum energy
in the ADD geometry. \\

\section{CASIMIR ENERGY IN RS GEOMETRY}

We now turn our attention to the RS model. Randall and Sundrum
introduce the following metric in a 5D brane world,
\begin{equation}
ds^2 = e^{-2ka|y|} \eta_{\mu \nu} dx^{\mu} dx^{\nu} + a^2 dy^2\, ,
\end{equation}
where $\mu,\nu=0,1,2,3, -\pi < y < \pi$, and $a$ is the radius of
the orbitfold $\mathcal{S}^1/\mathcal{Z}_2$ in the 5th dimension
$y$. The hidden, or Planck, brane locates at $y=0$ while the
visible, or TeV, brane locates at $y=\pi$. As is well-known, the
Planck-SM hierarchy is bridged if $ka\sim
\mathcal{O}\left(10\right)$ so that the mass scale at $y=\pi$ is
suppressed by the warp factor,
\begin{equation}
\frac{M_{\rm SM}}{M_{\rm Pl}}\sim e^{-\pi ka}\sim 10^{-16}\, .
\end{equation}
It is customary to take $k\sim M_{\rm Pl}$. So in the RS model the
extra dimension size $a$ is only about 10 times the Planck length.
We follow the original RS construct where only the gravity sector
lives in the bulk while all other fields in the standard model are
confined on the TeV brane. For our purpose we impose supersymmetry
to the system. Supersymmetric RS model has been investigated by
various authors\cite{Altendorfer:2000,Gherghetta:2000}. Since we
are concentrating on the SUSY-breaking of the gravity sector, it
suffices the purpose to consider only the graviton and gravitino
fields in the action\cite{Gherghetta:2001}:
\begin{equation}
S=S_5+S_{y=0}+S_{y=\pi}\, ,
\end{equation}
where the action in the bulk is,
\begin{eqnarray}
&S_5&=\int d^4x\int dy
\sqrt{-g}\Big[-\frac{1}{2}M_5^3\Big(\mathcal{R}\nonumber \\
&&+i\bar{\Psi}^i_M\gamma^{MNP}D_N\Psi^i_P
-i\frac{3}{2}\sigma'\bar{\Psi}^i_M\gamma^{MN}\sigma_3^{ij}\Psi^j_N\Big)
\nonumber \\
&&-\Lambda_5\Big]\, ,
\end{eqnarray}
and that on the two branes are
\begin{equation}
S_{y=y^*}=\int d^4x\sqrt{-g}[\mathcal{L}_{y^*}-\Lambda_{4,y^*}]\,
.
\end{equation}
Here $\mathcal{R}$ is the 5D Ricci scalar and $\bar{\Psi}^i$ the
5D symplectic Majorana gravitino fields ($i=1,2$) whose
left-handed and right-handed components satisfy the condition
$\gamma_5\Psi_{L,R}=\pm \Psi_{L,R}$, respectively, and $\Psi_{\mu
L}(\Psi_{\mu R})$ are even (odd) under $\mathcal{Z}_2$-parity.
Supersymmetry automatically ensures the conditions
$\Lambda_5/k=\Lambda_{4,\pi}=-\Lambda_{4,0}$. From our motivation,
we further assume that these are all identically zero. At $y^*=0$
the mass scale is of the order Planck scale, $M_{\rm Pl}^2\sim
M_5^3/k$. With $k\sim M_{\rm Pl}$, we have $M_5\sim M_{\rm Pl}\sim
{\rm TeV}$. The effective mass scale on the $y^*=\pi$ brane is
then of the order $M_{\rm Pl}e^{-\pi ka}$. In the last term of the
bulk action, $\sigma'\equiv d(k|y|)/dy=\pm k$ for $y\geq 0$ and
$y< 0,$ respectively.

 The Kaluza-Klein (KK) decomposition and
the associated eigen-modes for bosons and fermions in the RS
geometry have been well studied in recent
years\cite{Goldberger:99,Flachi:2001,Gherghetta:2000}. Goldberger
and Wise\cite{Goldberger:99} first studied the behavior of bulk
scalar field in the RS model. Flachi et al.\cite{Flachi:2001}
investigated that for the bulk fermion field. Gherghetta-Pomarol
(GP)\cite{Gherghetta:2000} extended the study to supersymmetric
bulk fields in the RS AdS geometry.

The bulk gravitino satisfies the 5D Rarita-Schwinger equation,
\begin{equation}
\gamma^{MNP}D_N\Psi_P-\frac{3}{2}\sigma'\gamma^{MP}\Psi_P=0\, .
\end{equation}
Decomposing the 5D gravitino, $\Psi_M$, and the 5D supersymmetric
parameter, $\eta$, as
\begin{eqnarray}
\Psi_{\mu L,R}(x^{\mu},y)&=&\sum_n
\psi_{\mu L,R}^{(n)}(x^{\mu})f_{L,R}^{(n)}(y)\, , \nonumber \\
\Psi_{5 L,R}(x^{\mu},y)&=&\sum_n
\psi_{5 L,R}^{(n)}(x^{\mu})f_{5 L,R}^{(n)}(y)\, , \nonumber \\
\eta_{L,R}(x^{\mu},y)&=&\sum_n
\eta_{L,R}^{(n)}(x^{\mu})f_{L,R}^{(n)}(y)\, ,
\end{eqnarray}
GP solved the equation of motion and found the $y$-dependent
gravitino wavefunctions as
\begin{equation}
f_L^{(n)}=\frac{e^{3k|y|/2}}{N_n}\Big[J_2\Big(\frac{m_n}{ke^{-ky}}\Big)
+bY_2\Big(\frac{m_n}{ke^{-ky}}\Big)\Big],
\end{equation}
\begin{equation}
f_R^{(n)}=\frac{e^{3k|y|/2}}{N_n}\Big[J_1\Big(\frac{m_n}{ke^{-ky}}\Big)
+bY_1\Big(\frac{m_n}{ke^{-ky}}\Big)\Big],
\end{equation}
where $m_n$ is the 4D gravitino mass for the $n$th mode and the
coefficient $b$ satisfies the condition
\begin{equation}
b(m_n)=-\frac{J_1(m_n/k)}{Y_1(m_n/k)}=b(m_ne^{\pi ka})\, .
\end{equation}
Solving this equation, the 4D KK gravitino masses are found to be
\begin{equation}
m_n\simeq \Big( n+\frac{1}{4}\Big)\pi k e^{-\pi ka}\quad \quad
(n>0)\, .
\end{equation}
Note that for the massless zero mode ($n=0$), the wavefunction is
localized at the Planck brane:
\begin{equation}
f_L^{(0)}(y)=\frac{1}{\sqrt{N_0}}e^{-k|y|/2}\, .
\end{equation}
The behavior of the graviton modes are similar. In particular
Randall and Sundrum\cite{Randall:1999b} have shown that the
graviton zero mode is also localized at the Planck brane. The 4D
mass spectrum for the graviton is identical to that of the
gravitino.

Invoking opposite fermion chiralities on the two
boundaries\cite{Gherghetta:2001}, i.e.,
\begin{eqnarray}
\psi(0)&=& \gamma_5\psi(0)\, , \nonumber \\
\psi(\pi a)&=& -\gamma_5\psi(\pi a)\, ,
\end{eqnarray}
supersymmetry is thus broken on the TeV brane. Gherghetta and
Pomarol show that this would induce a mass shift for the massless
zero mode at the tree level which is doubly suppressed by the warp
factor,
\begin {equation}
m_{3/2}\simeq \sqrt{8}ke^{-2\pi ka}\sim \Big(\frac{M_{\rm
SM}}{M_{\rm Pl}}\Big)^2 M_{\rm Pl}\, .
\end{equation}
Physically, such a double suppression from the Planck scale by the
RS warp factor is due to the fact that the zero mode wavefunction
is localized on the Planck brane, which results in an extra warp
factor suppression from the TeV scale. On the other hand the
graviton, which satisfies the periodic, or untwisted, boundary
condition, whose zero mode would remain massless at the tree
level. Thus the zero mode gravitino-graviton mass split is $\Delta
m_0= m_{3/2}$.

We are now ready to look for the Casimir energy under
SUSY-breaking. Casimir energy in the RS geometry has been
investigated by several
authors\cite{Naylor:2002,Setare:2004,Saharian:2005}. First we
examine the form of the 4D Casimir energy density under the RS
geometry. This can be most easily obtained by introducing a new
variable $z=e^{ka|y|}/k$ and rewrite the RS metric in a form which
is conformally flat in 5D:
\begin{equation}
ds^2 = \frac{1}{k^2z^2}\eta_{MN} dx^{M} dx^{N}\, ,
\end{equation}
where $M,N=0,...,4$. In these conformal coordinates the
exponential warp factor disappears, but the separation between the
Planck and TeV branes become far apart, i.e.,
$z_{\pi}-z_0=k(e^{\pi ka}-1)$\cite{rattazzi:2006}.

Based on these conformal coordinates, one can readily transcribe
from what deduced in Sec.2 the generic scaling of the Casimir
energy in the RS model,
\begin{equation}
\rho^{(4)}_{\rm Casimir} \sim \mp e^{-4\pi ka}a^{-4}\, ,
\end{equation}
for graviton and gravitino fields, respectively. Thus the net
Casimir energy is identically zero if SUSY is preserved. Under
SUSY-breaking, the scaling becomes
\begin{equation}
\rho^{(4)}_{\rm Casimir} \sim e^{-4\pi ka} a^{-2} \Delta m^2 \, .
\end{equation}
Identifying $k\sim 10/a\sim M_{\rm Pl}$ and ignore the numerical
factors, the zero-mode contribution to the Casimir energy density
conforms with what our inverted hierarchy Ansatz implies:
\begin{equation}
\rho^{(4)}_{\rm Casimir,0} \sim \Big[\Big(\frac{M_{\rm SM}}{M_{\rm
Pl}}\Big)^2 M_{\rm Pl}\Big]^4\, \sim \rho_{\rm DE}.
\end{equation}

Unfortunately, unlike the zero mode the higher KK mode
wavefunctions are localized on the TeV brane instead. As such the
mass of the $n$th KK mode for gravitino under twisted boundary
condition is shifted to (for $m_n \ll k$ and $ka \gg 1$)
\begin{equation}
m_n\sim \Big(n+\frac{3}{4}\Big)\pi k e^{-\pi ka}\, ,
\end{equation}
while that for the corresponding graviton KK mode under the
untwisted boundary condition remains unchanged.
Therefore the
mass-split between graviton and gravitino for the $n$th mode is of
the order
\begin{equation}
\Delta m_n\sim  \frac{\pi}{2}k e^{-\pi ka}\sim M_{\rm SM}\sim {\rm
TeV}\, .
\end{equation}
This energy scale certainly overshadows that of the zero-mode, and
our hard-earned minute Casimir energy would be totally
overwhelmed.

\section{SUMMARY}
Recent observational evidence indicates that the dark energy may
actually be the cosmological constant. We argue that the numerical
coincidence between the SM-Planck hierarchy and the inverted SM-DE
hierarchy implies a deeper connection between the two. Invoking
this connection as our guidance, we investigate the possibility of
interpreting the Casimir energy density induced in a SUSY-breaking
brane world as the dark energy. We found that for ADD geometry
this scenario works if and only if the extra dimensionality is
$n=2$. For the RS geometry, we found that based on the GP
SUSY-breaking boundary condition the graviton-gravitino KK zero
mode contributes to a Casimir energy which conforms with the
inverted hierarchy structure. Unfortunately the higher KK modes
under the same construction would develop a mass shift at
$\sim$TeV scale, which overshadows the zero mode contribution. Our
attempt is therefore unsuccessful.

One possible way to ameliorate this difficulty is to device a
different mechanism for SUSY-breaking under which the higher KK
modes would acquire mass shifts that is at or above the Planck
scale and would thus be cut off. Conversely, if there is a way
where the higher KK mode mass shifts can also be suppressed to the
same level as that of the zero mode, then the total Casimir energy
contributed from all modes might still retain the same energy
scale. These possibilities will be further investigated.

\end{document}